\documentclass[twoside]{sfm}

\usepackage[T2A]{fontenc}
\usepackage[english]{babel}
\usepackage{lscape}
\usepackage{graphicx}
\usepackage{amssymb}
\usepackage{bibspacing}

\def\figurename{Figure}
\def\tablename{Table}
\newcommand{\titl}[3]
{

\begin{flushleft}
\baselineskip 14pt
{\sffamily\bfseries\Large #1\\[2.1mm]

}
\baselineskip 12pt
{\large #2\\[2.1mm]

}
{\baselineskip 11pt
\small\it #3\\[5mm]

}
\end{flushleft}

}

\newcommand{\abstre}[1]{
{

\small\noindent
{\bfseries Abstract} \hspace{5mm}
\baselineskip 10pt
\noindent
#1

}}


\begin{document}

\def\kms{km\,s$^{-1}$}
\def\bz{$\langle B_{\rm z} \rangle$}
\def\degr{^\circ}
\def\aaps{A\&AS}
\def\aap{A\&A}
\def\apjs{ApJS}
\def\apj{ApJ}
\def\rmxaa{Rev. Mexicana Astron. Astrofis.}
\def\mnras{MNRAS}
\def\actaa{Acta Astron.}
\newcommand{\Tef}{T$_{\rm eff}$~}
\newcommand{\Vt}{$V_t$}
\newcommand{\CC}{$^{12}$C/$^{13}$C~}
\newcommand{\CDC}{$^{12}$C/$^{13}$C~}

\pagebreak

\thispagestyle{titlehead}

\setcounter{section}{0}
\setcounter{figure}{0}
\setcounter{table}{0}

\newcommand\cir{BS\,Cir}
\newcommand\vir{CU\,Vir}
\newcommand\uma{CQ\,UMa}
\newcommand\ori{V901\,Ori}
\newcommand{\zav}[1]{\left(#1\right)}
\newcommand{\hzav}[1]{\left[#1\right]}
\newcommand{\szav}[1]{\left{#1\right}}
\newcommand{\teff}{$T_{\mathrm{eff}}$}
\newcommand{\oc}{$\textit{O-C}$}
\renewcommand{\thefootnote}{}

\markboth{Mikul\'a\v{s}ek Z. et al.}{Ap stars with variable periods}

\titl{Ap stars with variable periods}{Mikul\'a\v{s}ek Z.$^{1,2}$, Krti\v{c}ka J.$^1$, Jan\'{\i}k J.$^1$, Zejda M.$^1$,
Henry~G.\,W.$^3$, Paunzen E.$^1$, \v{Z}i\v{z}\v{n}ovsk\'y J.$^{\dag 4}$\footnote{The article is dedicated to one of its co-authors -- Dr. Jozef \v{Z}i\v{z}\v{n}ovsk\'y who passed away on 15 June 2013.}, Zverko J.$^5$}
{$^1$Department of Theoretical Physics and Astrophysics,
    Masaryk University, Kotl\'a\v{r}sk\'a 2, CZ\,611\,37, Brno, Czech Republic, email: {\tt mikulas@physics.muni.cz} \\
 $^2$Observatory and Planetarium of Johann Palisa, V\v SB --
    Technical University, Ostrava, Czech Republic\\
 $^3$Tennessee State University, Nashville, Tennessee, USA\\
 $^4$Astronomical Institute of Slovak Academy of Science,
            Tatransk\'a Lomnica, Slovak Republic\\
 $^5$Tatransk\'a Lomnica 133, Slovak Republic}

\abstre{The majority of magnetic chemically peculiar (mCP) stars exhibit
periodic light, magnetic, radio, and spectroscopic variations that can be
adequately modelled as a rigidly-rotating main-sequence star with persistent
surface structures. Nevertheless, there is a small sample of diverse mCP
stars whose rotation periods vary on timescales of decades while the shapes
of their phase curves remain unchanged. Alternating period increases and
decreases have been suspected in the hot CP stars CU Vir and V901 Ori,
while rotation in the moderately cool star BS Cir has been decelerating.
These examples bring new insight into this theoretically unpredicted
phenomenon. We discuss possible causes of such behaviour and propose that
dynamic interactions between a thin, outer, magnetically-confined envelope
braked by the stellar wind, and an inner faster-rotating stellar body
are able to explain the observed rotational variability.}

\baselineskip 12pt

\section{Introduction}

The magnetic chemically peculiar (mCP) stars with abnormal surface chemical
composition and strong, global magnetic fields are the most suitable test
beds for studying rotational evolution in upper (B2 to F6) main sequence (MS)
stars. The overabundant elements in their atmospheres are concentrated into
large spot regions that persist for decades to centuries. As an mCP star
rotates, periodic variations in its brightness, spectrum, and magnetic field
are observed. Combining both new and archival observations of mCP stars
collected over the past several decades, we can reconstruct their period
evolution (if any) with high accuracy.

\section{Period changes due to stellar evolution}\label{evol}

The abnormal chemical abundances, spottedness, and strong global magnetic
fields have no influence on the inner structure of mCP stars, which are
evolving as regular upper-main-sequence stars. For such stars that
are mildly rotating and without significant angular momentum loss
(Meynet \& Maeder \cite{mam}), evolutional models predict their moments
of inertia $J(t)$ and their rotational periods $P(t)$ should change
roughly according to the simple relations:
\begin{equation}\label{CP}
J(t) \doteq J_0\,\exp\zav{\frac{t}{\tau_{\rm{MS}}}}
\ \Rightarrow\ \frac{\dot{P}}{P}=\frac{\dot{J}}{J} =\tau_{\mathrm{MS}}^{-1},\ \Rightarrow\ P(t) \doteq P_0\,\exp\zav{\frac{t}{\tau_{\rm{MS}}}},
\end{equation}
where $\tau_{\mathrm{MS}}$ is the MS duration of an individual star, and
$P_0$ and $J_0$ are the ZAMS values of rotational period and moment of
inertia. These relations predict that we should observe slowing of these
stars' rotation.

Even for the most massive He-strong mCP stars with MS lifetimes
$\tau_{\mathrm{MS}}=30$\,Myr, we estimate that
$\dot{P}\leq9\times10^{-11}\,\mathrm{d}^{-1}\,P$. For the B2Vp He-strong
star \ori\ (HD\,37776) with the period of $P=1.538$\,d we obtain
$\dot{P}\simeq1.4\times10^{-10}$. Even though \ori\ is one of the best
monitored mCP stars, the uncertainty in its value of
$\delta\dot{P}=5\times10^{-10}$ is several times larger than the expected
evolutionary changes. Consequently, the rotational slowing driven by
stellar evolution of MS stars is too small to measure with current
techniques. If evolutionary changes are the only driver of rotational
evolution, we would expect rotational periods in mCP stars to be
quite constant.

\section{mCP stars with variable periods}

Careful period analyses of several dozen mCP stars have been done to date.
These analyses confirm the expectations that CP star periods are, in
general, stable within the uncertainties of their measurement. However,
there is a small subgroup of mCP stars whose light curve shapes and
spectroscopic variability remain stable, but whose rotational periods are
variable (Pyper et al. \cite{pyper97,pyper13}, Mikul\'a\v{s}ek et al.
\cite{mik901,unstead,mik11}, Townsend et al. \cite{town}).

Here we examine period changes in three of the best-observed mCP stars --
\ori, \vir, and \cir.

\subsection{He-strong V901 Orionis}

\begin{figure}[!t]
\begin{center}
\hbox{\includegraphics[width=10.5cm]{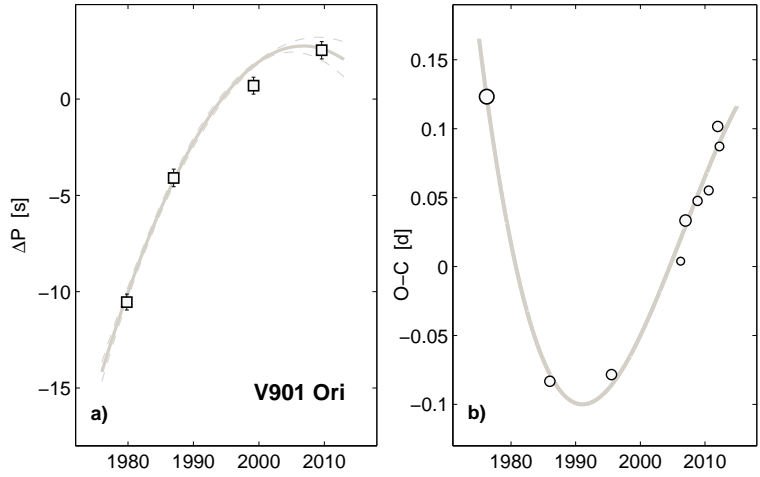}}
\vspace{-5mm}
\caption[]{a) Changes in the rotational period of \ori\ in seconds. Time
dependence of the period is approximated by a parabola reaching its maximum
in 2007. b) Changes in times of zero phase in days can be well fitted
by a cubic parabola.}
\label{duo901}
\end{center}
\end{figure}

\ori\ = HD 37776 is a very young He-strong B2p mCP star located in the
emission nebula IC 432 and has a strong $(B_s\approx20$~kG), complex,
global magnetic field (Thompson \& Landstreet \cite{thomla}, Kochukhov
et al. \cite{ko901}). Observed moderate light variations are caused by
spot regions of overabundant silicon and helium (Krti\v{c}ka et al.
\cite{krt901}).

Precise photometric and spectroscopic observations enabled Mikul\'a\v{s}ek
et al. \cite{mik07,mik901} to see continuous rotational deceleration,
increasing the period of $1.5387$\,d by a remarkable 18 s during the last
37 years! The deceleration was interpreted in terms of the rotational braking
of the outer stellar layers caused by angular momentum loss through the
stellar magnetosphere.  A complication with this interpretation, noted by
Mikul\'a\v{s}ek et al. \cite{mik901}, was the negative value of the period's
second derivative: $\ddot{P}=-29(13)\times 10^{-13}\,\mathrm{d}^{-1}$, which
implied that the rotational braking could soon change to acceleration.

Presently, we have about 3500 photometric and spectroscopic measurements
of \ori, spanning nearly four decades. We find $\overline{\dot{P}}=1.77(5)\times10^{-8},\ \tau_{\mathrm{spin}}= P/\overline{\dot{P}}=2.38(7)\times10^{5}\,\mathrm{yr}$ (only 1\% of $\tau_{\rm {MS}}$). Our current rate of the period
increase is $\ddot{P}=-32(4)\times10^{-13}\,\mathrm{d}^{-1}$, implying that
the deceleration of the stellar rotation had already switched to acceleration
by $2007\pm2$\,yr (Fig.\,\ref{duo901}a).

\subsection{Silicon mCP Star CU\,Virginis}

The rapidly-rotating silicon mCP star \vir\ (HD\,124224,  HR\,5313), with a
period of 0.5207\,d, displays more complex period changes. It is a hot
silicon mCP star with a mass and radius of 3\,M$_{\odot}$ and
2\,R$_{\odot}$, respectively (St\c{e}pie\'n \cite{step}) and
\teff$=13\,000$\,K, $\log g=4.0$ (Kuschnig et al. \cite{kusch}).
\vir\ is the only known MS star that shows variable radio emission,
resembling the radio lighthouse of pulsars (Trigilio et al. \cite{trigi00}).
It also displays strong variations in its brightness and the spectral lines
of He\,I, Si\,II, H\,I, and other ions. The nature of its light variability
in UV and optical regions was studied in detail by Krti\v{c}ka et al.,
\cite{krtcu}.

Occasional rapid increases in its rotation period have been reported.
Pyper et al. \cite{pyper97,pyper98} discovered an abrupt increase in
the period from $0.5206778$ to $0.52070854$\,d that occurred around
1984. Pyper et al. \cite{pyper13} recently discussed period changes in
the star based on their 2820 precise Str\"omgren $\mathit{uvby}$
values obtained with the Four College Automated Photometric Telescope
(FCAPT) between 1998 and 2012. They found the \oc\ values from
1993 forward implied a constant period of 0.5207137\,d, which is
the longest of its reported periods.

%

Mikul\'a\v{s}ek et al. \cite{mik11} collected and analyzed all available
observations of \vir\ containing phase information between 1949 and 2011.
They demonstrated that the shape of the phase curve was constant over
several decades while the period was continuously changing. The rotation
period gradually shortened until the year 1968 when it reached its minimum.
The period then started to lengthen, reaching a maximum at the end of their
data set. Much smaller stochastic-like period changes on a timescale of
several years were also reported.

\begin{figure}[!t]
\begin{center}
\hbox{\includegraphics[width=10.5cm]{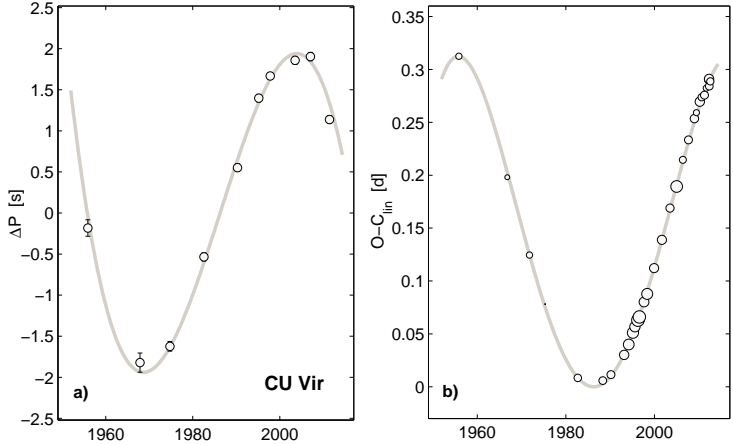}}
\vspace{-5mm}
\caption[]{a)~Changes in the rotation period of \vir\ in seconds with
respect to the mean $\overline{P}=0.52069383$\,d. The period variations
are approximated by an asymmetric cubic parabola that reaches a
minimum in 1969 and a maximum in 2004. The maximum
$\dot{P}=0.165\,\mathrm{s\,yr}^{-1}$ occurs in 1986. The amplitude of
the period variations is 3.9 s. b)~Changes in times of zero phase in
days can be well fitted by a biquadratic parabola. Each symbol represents
the mean of about 580 measurements with the the area proportional to the
weight of this set of observations.}
\label{duoCU}
\end{center}
\end{figure}

This is confirmed by our time series analysis also based on the recently
published FCAPT photometry from 1998--2012 (Pyper et al. \cite{pyper13}) as
well as our own measurements from 2011--2013. In total, we have 17\,936
individual measurements of \vir\, including 17\,241 photometric measurements
in passbands from 200 to 753 nm as well as spectroscopic, magnetic, and
radio observations. The period variations $P(t)$ can be well approximated
by the simple antisymmetric cubic parabola (see Fig.\,\ref{duoCU}a). We
assert that $P(t)$ reached its local minimum in the year 1968.7,
$P_{\mathrm{min}}=0.52067138$\,d, and its local maximum in the year
2003.9: $P_{\mathrm{max}}=0.52071628$\,d; we find the range of observed
periods to be 3.9 s. The rotational deceleration rate reached a maximum
of $\dot{P}=5.5\times10^{-9}=0.165\,\mathrm{s\,yr}^{-1}$.

We can quantify the deceleration rate using the spin-down time, $\tau$,
defined as $\tau=P(t)/\bar{\dot{P}}$, where $P(t)$ is the instantaneous
rotation period at the time $t$ and $\bar{\dot{P}}$ is the mean rate of
rotational deceleration. The paradox of \vir\ is, according to
Mikul\'a\v{s}ek et al. \cite{unstead}, that its spin-down time,
$ \tau \sim 6\times 10^5$ years, is more than two orders of magnitude
shorter than the estimated age of the star -- $9\times10^7$ yr
(Kochukhov \& Bagnulo \cite{kobacu}). For details see section \ref{nature}

\subsection{SrCrEu mCP star BS Circinis}

\begin{figure}[!t]
\begin{center}
\centering\includegraphics[width=7.8cm,angle=0]{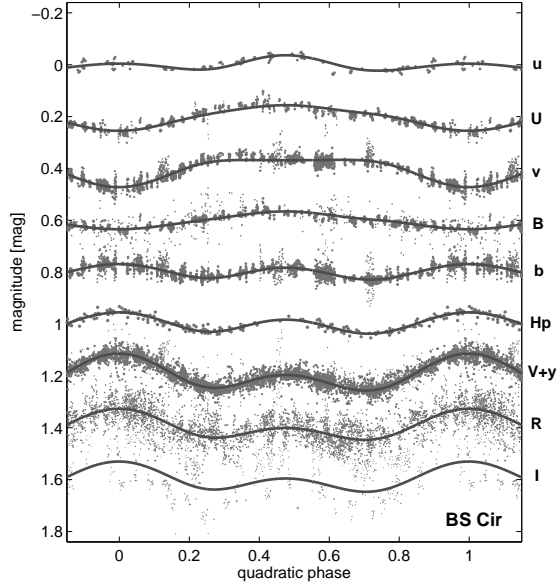}
\caption{Light curves of \cir\ plotted versus the phase calculated by means the quadratic ephemeris. The areas of individual symbols
are inversely proportional to their uncertainty, marked also by error bars. The observed light curves can be modelled if we assume two photometric spots of different colors centred at phases -0.001 and 0.475.}\label{crvBS}
\end{center}
\end{figure}
Both V901 Ori and CU Vir are among the hotter and thus more massive and
younger mCP stars.  To round out our sample of mCP stars, we included
several moderately cool SrCrEu mCP stars in our sample.  These stars
have relatively large photometric amplitudes, especially in the
Str\"omgren $v$ and $y$ passbands.  We find these cooler SrCrEu mCP stars
(BS Cir, CQ UMa, CS Vir, and VV Scl) all have stable periods with
the exception of \cir\ = HD\,125630 = HIP\,70346. Combining all available
kinematic, photometric, and spectroscopic data on \cir\, we derive the
following astrophysical parameters:
$T_{\mathrm{eff}}=8800\pm500\,\mathrm{K},\
L=41.7\pm1.4\,\mathrm{L}_{\odot},\ M=2.32\pm0.14\,\mathrm{M}_{\odot},\
\mathrm{age}= 510^{+90}_{-150}$\,Myr.
The star has a bipolar magnetic field with $B_{\rm{p}}$ of several kG
(Kochukhov \& Bagnulo \cite{kobacu} and Hubrig et al. \cite{hub06}).

\begin{figure}[!t]
\begin{center}
\centering\includegraphics[width=9.8cm,angle=0]{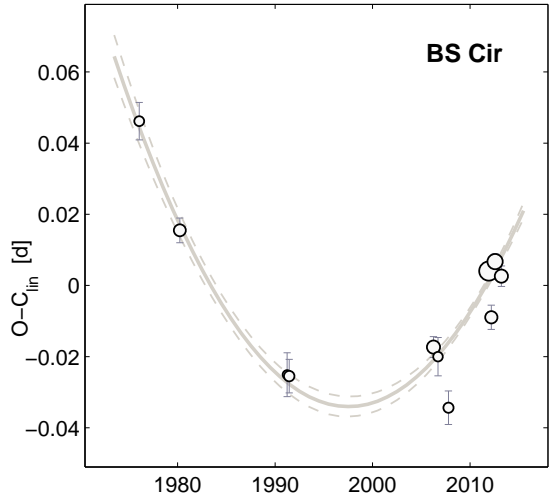}
\caption{The \oc\ curve of \cir\ can be well fitted by a parabola. The
period is lengthening linearly, $\dot{P}=5.6(4)\times10^{-9}$, while the
mean period is $\overline{P}=2.204285$\,d. The 1-$\sigma$ uncertainties
from the fit are denoted by dashed lines. The areas of individual symbols
are inversely proportional to their uncertainty, marked also by error
bars.}\label{OCBS}
\end{center}
\end{figure}

Our period analysis of \cir\ is based on all available observational material,
including 14\,488 individual photometric measurements in 11 data sets
that cover, more or less evenly, a time interval of 38 years. The weighted
scatter in these data of one measurement is 0.017 mag. \cir\ is among the
best photometrically observed mCP stars. The observed light curves were
acquired in 9 filter passbands and show very disparate shapes and amplitudes
at various effective wavelengths. Fortunately, all of the light curves can
be easy modelled if we assume two different photometric spots centred at
phases $-0.0014\pm0.0005$ and $0.4752\pm0.0010$, which are likely to be the
phases of the magnetic field positive and negative extrema.

We determine the mean rotational period and its time derivative to be
$\overline{P}=2.2042850(6)\,\mathrm d,\ \dot{P}=5.6(4)\times
10^{-9}=0.181(13)\,\mathrm{s\,yr}^{-1}$, respectively. The rate of period
increase is well determined: $\dot{P}/\delta{\dot{P}}>13.7$. The spin-down
time is $\tau_{\mathrm{spin}}=P/\dot{P}=1.05(8)$\,Myr, that is 0.2\% of the
estimated age of the star. Our method for determining the spin-down rate of
\cir\ is able to distinguish spin-down times seven times longer.

\section{Nature of the observed period changes}\label{nature}

There are several standard explanations for period variations in mCP stars,
all of which assume that mCP stars rotate as solid bodies: (1) changes
in the radius and mass distribution during MS evolution, (2) changes
due to angular momentum loss via a standard stellar wind, (3) changes caused
by angular momentum loss via a magnetized stellar wind, (4) precession of
the rotational axis, and (5) light-time effects caused by an undetected
companion in a binary system.

The first three mechanisms are expected to cause lengthening of the rotational
periods at a constant rate, mathematically: $\dot{P}>0,\ \ddot{P}=0$. The
spin-down time of evolutionary period changes (30 Myr minimum) is always
at least three time larger than the present limit of detection -- 10 Myr.
The spin-down time caused by a standard stellar wind is several orders
larger than the evolutionary timescale and, therefore, completely
undetectable. Angular momentum loss via a magnetized stellar wind is
detectable only in the extremely hot mCP stars with very strong magnetic
fields $(B_{\mathrm p}>5$\,kG). The only known mCP star for which the observed
spin-down time of 1.34 Myr (Townsend et al. \cite{town}) agrees with
theoretical predictions is very rare hot Be+He-strong hybrid object
$\sigma$\,Ori\,E.

The fourth mechanism, precession of the rotational axis, should manifest by
cyclic changes in the shape of the light curves, but this is not observed.
In addition, the amplitude of precessionally-induced period changes are
generally negligible (Mikul\'a\v{s}ek et al. \cite{mik901}). Finally, the
fifth mechanism, cyclic changes in the observed period due to the light-time
effect from an invisible companion, should be accompanied by radial
velocity variations, but radial velocity changes are not observed (Pyper et
al. \cite{pyper98}, Mikul\'a\v{s}ek et al. \cite{mik901}).

The discovery of variable rotation in the cooler SrCrEu mCP star BS Cir shows
that this phenomenon may occur across all mCP types and maybe all upper
MS stars.

Therefore, we ``abandon the assumption of the rigid rotation,'' as suggested by
St\c{e}pie\'n \cite{step}, and examine the alternative concept that the
structure of the surface layers of mCP stars are dominated by global magnetic
field and can rotate differentially from the denser interior of a star.
The magnetic field contributes to immobilization of the outer parts of mCP
stars in the vertical direction and also prevents the spot structures from
dissolving in the horizontal direction.  The magnetic field could
accomplish this only if its energy density is larger than the energy of
the stellar dissipative motions. Simple calculations show that, in the
photospheres of A and B-type stars, even very weak magnetic field is
sufficient.


If the magnetic fields of upper MS stars are fossil fields, then all
photospheres should be, to some extent, controlled by magnetic field.
Consequently, transient spot structures on moderately hot `normal' (non-CP)
MS stars are possible. The depth of magnetic field dominance (the thickness
and endurance of spot structures) strongly depends on the strength of the
magnetic field. Such an `eggshell' surface layer(s) dominated by magnetic
field may behave as a solid body. Even weak interaction with the outer
environment or the stellar interior may be able to accelerate or decelerate
this layer very effectively.

These considerations and speculations are a challenge for theoreticians
modelling the structure of upper MS stars.
\bigskip

{\it Acknowledgements.} This work was funded by the grant of GA\v{C}R P209/12/0217 and SoMoPro (3SGA5916).

\end{document}